\documentclass{article}
\usepackage{authblk}
\usepackage{amsmath,graphicx,amssymb,mathrsfs}
\usepackage{amsfonts}
\usepackage{color}
\usepackage{subcaption}
\usepackage{authblk}
\usepackage[margin=1in]{geometry}
\usepackage{algorithm}% 
\usepackage{lineno}
\usepackage{makecell}
\DeclareRobustCommand{\vect}[1]{
	\ifcat#1\relax
	\boldsymbol{#1}
	\else
	\mathbf{#1}
	\fi}

\makeindex

\begin{document}
\title{A new computational model for quantifying blood flow dynamics across myogenically-active cerebral arterial networks}

\author[1,2*]{Alberto Coccarelli}
\author[1]{Ioannis Polydoros}
\author[1]{Alex Drysdale}
\author[3]{Osama F. Harraz}
\author[4]{Chennakesava Kadapa}

\affil[1]{Zienkiewicz Institute for Modelling, Data and AI, Faculty of Science and Engineering, Swansea University, Swansea, UK}
\affil[2]{Department of Mechanical Engineering, Faculty of Science and Engineering, Swansea University, Swansea, UK}

\affil[3]{Department of Pharmacology, Larner College of Medicine, and Vermont Center for Cardiovascular and Brain Health, University of
Vermont, Burlington, USA}
\affil[4]{
School of Computing, Engineering and the Built Environment,
 Edinburgh Napier University, Edinburgh, UK}

\affil[*]{corresponding author; e-mail: alberto.coccarelli@swansea.ac.uk; address: Engineering North 321, Swansea University Bay Campus, Fabian Way, Crymlyn Burrows, Skewen, Swansea SA1 8EN}

\maketitle
\begin{abstract}
Cerebral autoregulation plays a key physiological role by limiting blood flow changes in the face of pressure fluctuations. Although the involved cellular processes are mechanically driven, the quantification of haemodynamic forces in in-vivo settings remains extremely difficult and uncertain.
%Cerebral autoregulation is essential for healthy brain function and recent technological advances enabled to appreciate several features of the involved processes under in-vivo conditions.
In this work, we propose a novel computational framework for evaluating the blood flow dynamics across networks of myogenically active cerebral arteries, which can modulate their muscular tone to stabilize flow (and perfusion pressure) as well as to limit vascular intramural stress.
The introduced framework is built on contractile (myogenically active) vascular wall mechanics and blood flow dynamics models, which can be numerically coupled in either a weak or strong way. We investigate the time dependency of the vascular wall response to pressure changes at both single vessel and network levels. The robustness of the model was assessed by considering different types of inlet signals and numerical settings in an idealized vascular network formed by a middle cerebral artery and its three generations. For the vessel size and boundary conditions considered, weak coupling ensured accurate results with a lower computational cost. To complete the analysis, we evaluated the effect of an upstream pressure surge on the haemodynamics of the vascular network. This provided a clear quantitative picture of how pressure and flow are re-distributed across each vessel generation upon inlet pressure changes. This work paves the way for future combined experimental-computational studies aiming to decipher cerebral autoregulation.
\end{abstract}

\section{Introduction}
%Journal suggestion (by Chenna): \url{https://www.sciencedirect.com/journal/computer-methods-and-programs-in-biomedicine}

Due to their size and extension, small arteries and arterioles are responsible for a significant blood pressure drop across the cerebral circulation~\cite{blanco2017}. Diameter in these resistance vessels is regulated through a combination of local and systemic control mechanisms that operate muscular apparatus made of smooth muscle cells (SMCs)~\cite{claassen2021}. This enables blood vessels to develop tone across their wall thickness and ultimately to adjust their inner diameter upon different mechanical stimuli. From a hierarchical point of view, the myogenic tone serves as a fundamental, low-level mechanism controlling lumen diameter, as it directly responds to the local pressure level. This mechanism enables the vessel to limit diameter fluctuations and stabilize flow in the face of significant hydrodynamic changes by developing tone within its muscle layer. This means that, in response to an increase in upstream pressure, the vessel will ultimately reduce its diameter in an attempt to maintain a relatively constant flow rate.  Haemodynamic forces, such as luminal pressure and shear stress, are sensed in different ways across vascular compartments, having a different impact on vascular contractility~\cite{knot1998,harraz2022,klug2023}.
The presence of basal vascular tone is also essential for flow and metabolic control, which operates by relaxing the local vasculature to divert blood flow towards regions where it is needed the most~\cite{longden2017}. From an experimental perspective, isolating these regulatory mechanisms is a cumbersome task, and due to their inter-dependency, the distribution of mechanical forces along the considered vascular segments remains extremely uncertain. Given the key role of mechanical stimuli in cerebral vascular function, developing methods for their systematic quantification is urgently needed.

Computational blood flow dynamics, in conjunction with biologically-motivated vascular wall models, can be used to shed light on different aspects of cerebral autoregulation. Diverse methodologies were proposed to describe the vascular response to acute changes in haemodynamic forces. Whilst the seminal work by Carlson et al.~\cite{carlson2008} introduced a general framework which integrates together myogenic, flow and metabolic controls without directly including sub-tissue scales, many other authors, including us, have adopted multi-scale approaches for modelling specific components of the contractile machinery in muscular arteries~\cite{yang2003a,yang2003b,coccarelli2018,uhlmann2023,coccarelli2024}. Several studies used a high-level description of underlying regulatory processes to investigate the impact of autoregulation on cerebral blood flow within realistic vessel network models~\cite{aletti2016a,aletti2016b,tong2021,daher2023a}. 
However, in experimental settings, vascular control mechanisms are typically characterized by using a broad repertoire of compounds that selectively activate or inhibit specific cellular components. To elucidate the causative links between drug intervention, luminal pressure and vessel wall deformation, it is essential to develop multi-scale modelling methodologies that can mimic the effect of intracellular biochemical processes on the tissue emergent behaviour.
%what we are proposing
In this respect, we have recently introduced a vascular mechanics model~\cite{coccarelli2024} able to recapitulate all the major pressure-induced intracellular (Ca$^{2+}$, ROCK and PKC) pathways in cerebral arteries, which translates the mechanical stimulus into SMC contractility, which in turn is integrated into an emergent tissue response.
In the current study, we introduce a methodology for integrating this multi-scale mechano-biological model within an extensively validated blood flow dynamics framework. This provides an in-silico tool to predict and recover various hemodynamic and vascular wall quantities under a wide range of conditions, such as variable upstream or downstream pressure or the presence of vasoactive agents, that can be replicated in the laboratory.

\section{Methods}
\subsection{Blood flow dynamics}\label{bfd}
We assume that flow in small cerebral arteries is laminar and axisymmetric, with a Poiseuille velocity profile. The considered 1-D fluid domain may range from a single vessel to a complex vessel network. The mass and momentum conservation equations for a fluid flowing in a collapsible vessel can be written in the pressure-flow form as~\cite{carson2017,coccarelli2021}:
\begin{equation}\label{flowCons}
\begin{cases}
C_A\frac{\partial P}{\partial t}+\frac{\partial Q}{\partial z}=0,  \\
\frac{\rho}{A}\frac{\partial Q}{\partial t}+\frac{\rho}{A} \frac{\partial}{\partial z}(\frac{Q^2}{A})+ \frac{\partial P}{\partial z} +8\pi\mu\frac{Q}{A^2}=0,
\end{cases}
\end{equation}
where $z$ is the axial direction, $A$ is the luminal cross-sectional area, $P$ is the average pressure in the cross-section corresponding to pressure acting on the inner wall surface, $Q$ is the volumetric flow rate in the cross-section whilst $\rho$  and $\mu$ are respectively the blood density and dynamic viscosity which, for the sake of simplicity, are assumed constant. It is worth mentioning that the variation of the area with respect to the fluid pressure defines the vessel compliance $C_A=\frac{\partial A}{\partial P}$ and can be determined from the constitutive law of the vascular wall. In line with~\cite{carson2017}, the system (\ref{flowCons}) is linearized with respect to time as follows
\begin{equation}\label{conslin}
\begin{cases}
C_A^n\frac{\partial P^{n+1}}{\partial t}+\frac{\partial Q^{n+1}}{\partial z}=0,  \\
\frac{\rho}{A^n}\frac{\partial Q^{n+1}}{\partial t}+\frac{\partial P^{n+1}}{\partial z} =-\Bigg(\frac{\rho}{A}\frac{\partial}{\partial z}(\frac{Q^2}{A})+8\pi\mu\frac{Q}{A^2}\Bigg)^n,
\end{cases}
\end{equation}
where $n+1$ and $n$ represent the current and previous time steps. The fluid domain is subdivided into elements of non-necessarily equal size. Following the work by Carson and Van Loon~\cite{carson2017}, Eqs. (\ref{conslin}) are integrated in space using the enhanced trapezoidal rule method and discretized in time using a second-order backward difference scheme. After some steps, the system of equations in (\ref{conslin}) may be re-written at the element level in the following compact form:
\begin{equation}\label{compact}
\textbf{F}_e \textbf{P}_e^{n+1}+\textbf{G}_e \textbf{Q}_e^{n+1}=\textbf{h}_e^{n},
\end{equation}
in which $e$ represents the elemental level, $\textbf{F}_e$, $\textbf{G}_e$ are the stiffness matrices of pressure and flow, $\textbf{P}_e^{n+1}$ and $\textbf{Q}_e^{n+1}$ the vectors containing the current element values of pressure and flow, and $\textbf{h}_e$ the vector representing convection and diffusion components evaluated at previous time step. 
Eqs. (\ref{compact}) serves for the assembling of the global system matrix, which, in conjunction with the boundary conditions, are used to compute (through `spsolve' from SciPy 1.6.0) the pressure and flow rate to the next time step ($P^{n+1}$, $Q^{n+1}$) across the whole fluid domain. Bifurcations, unifications, vessel geometry, and material discontinuities are handled in line with Carson and Van Loon~\cite{carson2017}. The luminal cross-sectional area and wall compliance at the next time step ($A^{n+1}$, $C_A^{n+1}$) are subsequently recovered from the vascular wall constitutive relationship, as reported in the following.

\subsection{Multi-scale vascular mechanics}\label{msvm}
The mechanics across the vascular wall are described by following the bottom-up approach reported in our previous work~\cite{coccarelli2024}. The wall of cerebral vessels is a complex structure endowed with the capacity to generate tone upon pressure loading. Although the wall is made of functionally different layers, we assume that the totality of its volume is occupied by SMCs. SMCs' contractile activity is described through two submodels, which describe i) the pressure-induced biochemical signalling and ii) the myosin-actin interaction alongside cytoskeleton remodelling. 

\subsubsection{Intracellular chemo-mechanics}
The intracellular pathways activated by pressure are represented through the following (normalized) quantities: Ca$^{2+}$ concentration ($\xi_0$), ROCK activity level ($\xi_1$), HSP27 phosphorylation level ($\xi_2$), MLCP phosphorylation level ($\xi_3$), Cofilin phosphorylation level ($\xi_4$), LC$_{20}$ phosphorylation level ($\xi_5$), and G-actin content ($\xi_6$). The time evolution of these variables is evaluated through a logic-based signalling graph, which is translated in the following system of equations:

\begin{equation}\label{xidyn}
\begin{aligned}
	\frac{d \xi_0}{dt}&=\frac{1}{\tau_{\text{c}0}}(\chi_0-\xi_0),\\
 \frac{d \xi_1}{dt}&=\frac{1}{\tau_{\text{c}1}}(\chi_1-\xi_1),\\
 \frac{d \xi_2}{dt}&=\frac{1}{\tau_{\text{c}2}}(\chi_2-\xi_2),\\
 	\frac{d \xi_3}{dt}&=\frac{1}{\tau_{\text{c}3}}[(1-\chi_3)-\xi_3],\\
  	\frac{d \xi_4}{dt}&=\frac{1}{\tau_{\text{c}4}}(\chi_4-\xi_4),\\
   	\frac{d \xi_5}{dt}&=\frac{1}{\tau_{\text{c}5}}[ \chi_5(1-\xi_5)-(1-\chi_6)\xi_5],\\
 	\frac{d \xi_6}{dt}&=\frac{1}{\tau_{\text{c}6}}[(1-\chi_7)+(1-\chi_8)-(1-\chi_7)(1-\chi_8)-\xi_6].
 \end{aligned}
\end{equation}
where $\chi_i$ with $i$=0,$\ldots$,8 is a logistic function connecting two signalling variables whilst $\tau_{\text{c}j}$ with $j$ = 0,$\ldots$,6 represent the time constants associated to each intracellular process. Eqs. (\ref{xidyn}) allow us to evaluate how changes in pressure (over time) influence cross-bridges (XBs) formation (between actin and myosin) and cytoskeleton remodelling, which are represented by $\xi_5$ and $\xi_6$, respectively. These molecular factors and SMC stretch level drive the relative sliding between actin and myosin filaments, which, together with cytoskeleton stiffness (represented by the F-actin content), enable tone development. Given a luminal pressure level ($P$), Eqs. (\ref{xidyn}) are discretized in time with the two-step Adams-Bashforth method and solved through a non-linear equations solver (`root' from SciPy 1.6.0, method=`lm’). For the sake of simplicity, all the SMCs are assumed to be aligned along the circumferential direction, and only the corresponding stretch component ($\lambda_{\theta}$) plays a significant role in tone development. The cell contractile fibres (CFs) can be represented as a series of interconnected contractile units (CUs) that are anchored at the cell membrane through actin cortex passive elements (see ~\cite{murtada2016} for more details). The dynamics governing the (normalized) relative filament sliding $(\bar{u}_{\text{fs}})$ within each CU is expressed via
\begin{equation}\label{dufs}
	\frac{d \bar{u}_{\text{fs}}}{dt}=\frac{1}{\tau_{\text{m}}}(F_{\text{a}}-F_{\text{c}}) +\frac{1}{2N_{\text{CU}}}\frac{d \lambda_{\theta}}{dt},
\end{equation}
where $\tau_{\text{m}}$ is the time constant associated with actin-myosin filaments sliding dynamics (and force generation). $F_{\text{c}}$ is the average driving force generated from the XBs cycling
\begin{equation}
	F_{\text{c}}=\bar{L}_{\text{fo}}\:\frac{L_\text{m}}{\delta_\text{m}}\:\xi_5\:n_{\text{XBmax}} \:k_{\text{XB}}\:u_{\text{PS}},
\end{equation}
where $L_\text{m}$ is the average length of myosin filaments, $n_{\text{XBmax}}$ is the maximum phosphorylation rate, $\delta_{\text{m}}$ is the average distance between myosin monomers heads, $k_{\text{XB}}$ represents the XB elastic stiffness and $u_{\text{PS}}$ is the average displacement associated to power-stroke, whilst $\bar{L}_{\text{fo}}$ describes the filament overlap, which depends on the relative filament sliding $\bar{u}_{\text{fs}}$ via
\begin{equation}
    \bar{L}_{\text{fo}}=\text{exp}[\frac{(\bar{u}_{\text{fs}}-\bar{u}_{\text{fs}}^{\text{opt}})^2}{2(s_{\text{f0}}/L_m)^2}].
\end{equation}
The reaction force due to the resistance from the number of contractile units ($N_{\text{CU}}$) in series with a F-actin element at each extremity is given as a product between the total (XBs and passive elements) elongation and the resulting stiffness of the contractile fibres:	
\begin{equation}\label{Fa}
F_{\text{a}}=(\lambda_{\theta}-1-2N_{\text{CU}}\bar{u}_{\text{fs}})\frac{k_{\text{tCU}}k_{\text{AC}}}{2k_{\text{tCU}}+k_{\text{AC}}}.
\end{equation}
where $k_{\text{tCU}}$ is the stiffness associated with a number of CUs ($N_{\text{CU}}$), which is directly related to the level of LC$_{20}$ phosphorylation via
\begin{equation}
k_{\text{tCU}}=\frac{\bar{L}_{\text{fo}}\:L_\text{m}\:\xi_5\:n_{\text{XBmax}}\:k_{\text{XB}}}{2\delta_{\text{m}}N_{\text{CU}}},
\end{equation}
whilst $k_{\text{AC}}$ is the actin cortex stiffness, which can be evaluated as a function of the F-actin content level
\begin{equation}	k_{\text{AC}}=k_{\text{ACmax}}\frac{\xi_7^{n_{\text{AC}}}}{\xi_7^{n_{\text{AC}}}+{K_{\text{AC}}}^{n_{\text{AC}}}},
\end{equation}
where $k_{\text{ACmax}}$ is the maximum stiffness under loading conditions, whilst $n_{\text{AC}}$ and $K_{\text{AC}}$ are the coefficients of the associated activation function. 

\subsubsection{Tissue mechanics}
The vessel wall consists of multiple concentric SMC layers, and here, in line with~\cite{coccarelli2021,coccarelli2023}, it is modelled as an axisymmetric homogeneous hyperelastic thick-walled tube. The stretch $\lambda_{\theta}$ experienced by an SMCs layer coincides with the circumferential stretch of that portion of tissue. Due to the variability of $\lambda_{\theta}$ along the wall thickness, it is possible to have (slightly) different intracellular signalling for SMCs located in different layers. Hence, we subdivide the wall thickness into a finite number of SMCs layers $n_{\text{CL}}$, each of them having intracellular chemo-mechanics which depends on the average $\lambda_{\theta}$ over the layer.
Finite strain theory and incompressibility assumption are adopted for describing the tissue kinematics
\begin{equation}
\lambda_r=\frac{R}{rk_{\omega}\lambda_z},~~~\lambda_\theta=\frac{k_{\omega}r}{R},~~~\lambda_r\lambda_\theta\lambda_z=1,
\end{equation}
where $\lambda_{r}$ and $\lambda_z$ are the radial and axial stretches, respectively, whilst $R$ is the reference radius, $r$ the deformed radius and $k_{\omega}$ a parameter accounting for the residual strain. The usual mapping between reference ($\Omega_R$) and deformed ($\Omega_D$) configuration is used:
\begin{equation}
\Omega_{\text{R}}\rightarrow \Omega_{\text{D}}:~~~r=\sqrt{\frac{(R_{\text{i}}+H)^2-R_{\text{i}}^2}{k_{\omega}\lambda_z}+r_{\text{i}}^2},
\end{equation}
where $r_{\text{i}}$ is the deformed luminal radius, whilst $R_{\text{i}}$ and $H$ are the luminal radius and thickness in the reference configuration, respectively. The luminal circumferential stretch is indicated with $\lambda_{\theta \text{i}}=\frac{k_{\omega}r_\text{i}}{R_\text{i}}$ while the deformed luminal and outer diameters are indicated by $d_{\text{i}}$ and $d_{\text{o}}$, respectively. The generated tone from CFs can be quantified through the average First Piola-Kirchhoff stress $P_{\text{a}}$ of the vascular tissue, $P_{\text{a}}=N_{\text{CF}}F_{\text{a}}$, where $N_{\text{CF}}$ is the surface density of the CFs.  
The momentum conservation principle along the radial direction allows us to link the luminal cross-sectional area ($A$ = $\pi r_{\text{i}}^2$ =$\pi \frac{\lambda_{\theta \text{i}}^2R_{\text{i}}^2}{k_{\omega}^2})$ to luminal pressure $P$~\cite{coccarelli2021}:
\begin{equation}\label{pres}
P=P_{\text{ext}}+\int_{R_{\text{i}}}^{R_{\text{i}}+H} (\lambda_{\theta}\frac{\partial \Psi}{\partial \lambda_{\theta}}-\lambda_{r}\frac{\partial \Psi}{\partial \lambda_{r}})\frac{dR}{\lambda_{\theta}\lambda_{z}r},
\end{equation}
where $P_{\text{ext}}$ is the external pressure acting on the outer surface of the vessel, $\Psi$ is the material {strain-energy function}, which is decomposed into active and passive components. The former can be evaluated as
\begin{equation}
\Psi_\text{a}=\int{P_\text{a}~d\lambda_{\theta}}=\frac{N_{\text{CF}}}{2}\frac{k_{\text{tCU}}k_{\text{AC}}}{2k_{\text{tCU}}+k_{\text{AC}}}(\lambda_{\theta}-1-2N_{\text{CU}}\bar{u}_{\text{fs}})^2,
\end{equation}
whilst the passive behaviour is described in line with~\cite{holzapfel2000}
\begin{equation}
\Psi_\text{p}=c_0(I_1-3)+\frac{c_1}{2c_2}\{\text{exp}[c_2(I_4-1)^2]-1\},
\end{equation}
where $c_0$, $c_1$ and $c_2$ are the media constitutive parameters, $I_1=\lambda_{r}^2+\lambda_{\theta}^2+\lambda_{z}^2$, $I_4=\lambda_{\theta}^2 \text{cos}^2\phi+\lambda_{z}^2\text{sin}^2\phi$ with $\phi$ being the orientation angle {of a collagen fibres family, which are oriented along the circumferential direction of the vessel. 

Whilst Eq. (\ref{pres}) concerns the momentum balance across the whole wall thickness, Eq. (\ref{dufs}) describes the time dependency of quantities that depend on the radial coordinate $r$. The integral in Eq. (\ref{pres}) depends on both $\lambda_{\theta}$ and  $\bar{u}_{\text{fs}}$ since $\frac{\partial \Psi_{{a}}}{\partial \lambda_{\theta}}=N_{\text{CF}}\frac{k_{\text{tCU}}k_{\text{AC}}}{2k_{\text{tCU}}+k_{\text{AC}}}(\lambda_{\theta}-1-2N_{\text{CU}}\bar{u}_{\text{fs}})$ and can be computed via Simpson's rule. In this study the adopted number of integration points (5) coincides with the number of cellular layers $n_{\text{CL}}$ and therefore $\lambda_{\theta}$ and  $\bar{u}_{\text{fs}}$ are computed at the middle of each cellular layer and represented by $\lambda_{\theta,k}$, $\bar{u}_{\text{fs},k}$ with $k$=1,$\ldots$,$n_{\text{CL}}$. To evaluate the mechanics across the vascular wall we rewrite Eqs. (\ref{dufs},\ref{pres}) as residuals
\begin{equation}\label{combstr}
\begin{aligned}
    \mathcal{F}_k(\bar{u}_{\text{fs},k},A) &= \frac{d \bar{u}_{\text{fs},k}}{dt} - \frac{1}{\tau_{\text{m}}}(F_{\text{a},k} - F_{\text{c},k}) - \frac{1}{2N_{\text{CU}}}\frac{d \lambda_{\theta,k}}{dt} = 0,~~~\text{with}~~~k=1,\ldots,n_{\text{CL}};\\
    \mathcal{J}(\bar{u}_{\text{fs}},A) &= P - P_{\text{ext}} - \int_{R_{\text{i}}}^{R_{\text{i}}+H} \left(\lambda_{\theta}\frac{\partial \Psi}{\partial \lambda_{\theta}} - \lambda_{r}\frac{\partial \Psi}{\partial \lambda_{r}}\right) \frac{dR}{\lambda_{\theta}\lambda_{z}r} = 0.
\end{aligned}
\end{equation}
The set of Eqs. (\ref{combstr}) is discretized in time with the two step Adams-Bashforth method and solved by using a solver for non-linear equations (`root' from SciPy 1.6.0, method=`lm') providing the solution ($\bar{u}_{\text{fs}}^{n+1},A^{n+1}$) at the next time step. Once the area $A^{n+1}$ is known, the compliance $C_A^{n+1}$ can be evaluated using a first-order centered finite difference scheme as done in our previous work~\cite{coccarelli2021,coccarelli2024}.
Alternatively, the vessel wall mechanics can be evaluated by assuming that the active stress component depends only on the average trans-mural $\bar{u}_{\text{fs}}$,  which is obtained by considering the average circumferential stretch over the vessel thickness. This strategy, already employed in~\cite{coccarelli2024}, is more computationally efficient than the original without `averaged active stress' as it requires simultaneously solving only two equations rather than the set of Eqs. (\ref{combstr}).

\subsection{Multi-physics coupling strategy}
The fluid and solid wall equations are strongly coupled via the fixed point iteration method (Fig. \ref{scheme}). The guess values for the cross-sectional area $A^k$ across the whole vascular network are initialized by considering the luminal areas at the previous time step. The numerical solution of 1-D fluid flow equations (reported in \ref{bfd}) provides the pressure and volumetric flow fields ($P^{k+1}$, $Q^{k+1}$) throughout the fluid domain. Each fluid node is associated with a vascular tissue ring, whose mechanics do not depend on the neighbouring nodes along the vessel's axial direction. The vascular chemo-mechanical model (reported in Section~\ref{msvm}) is used to evaluate sequentially the intracellular signalling variables $\xi_5$ and $\xi_6$ and the resulting luminal area $A^{k+1}$, alongside the relative filament sliding $\bar{u}_{\text{fs}}^{k+1}$.
In the current methodology Eqs. (\ref{xidyn}) only depend on the current luminal fluid pressure $P^{k+1}$ and, therefore, do not need to be solved simultaneously with Eqs. (\ref{combstr}). The new value for the cross-sectional area $A^{k+1}$ is used to update the wall compliance $C_A^{k+1}$. If the Root Mean Square Relative Error (RMSRE) between $A^{k+1}$ and $A^{k}$ across the vascular network is below the prescribed tolerance $\epsilon$, the variables $P^{k+1}$, $Q^{k+1}$, $A^{k+1}$, $\bar{u}_{\text{fs}}^{k+1}$ are stored as the solution of the current time step; otherwise the fluid-structure interaction sequence is again re-iterated with $A^{k+1}$ as the new guess values.
\begin{figure}
\centering
\includegraphics[]{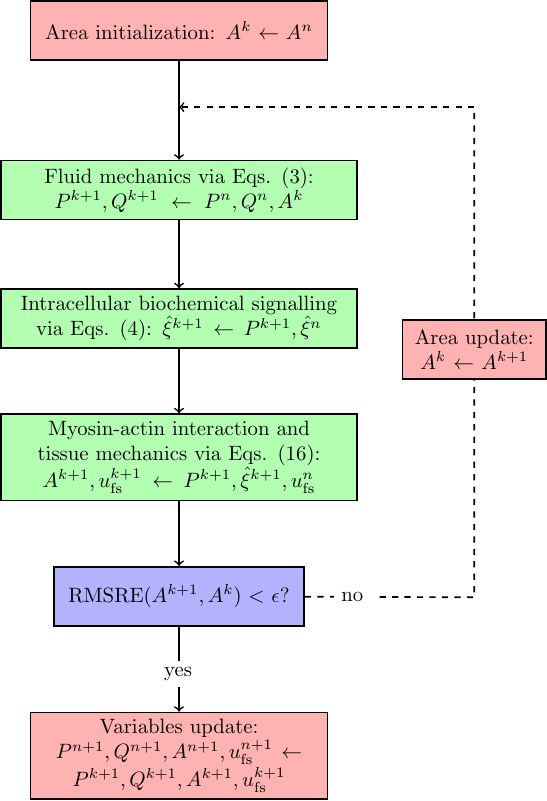}
\caption{Model components coupling. Whilst Eqs. (3) are simultaneously solved for all vascular network nodes, Eqs. (4) and (16) are solved independently for each vascular network node.}
\label{scheme}
\end{figure}
The tolerance for the fluid-structure interaction coupling ($\epsilon$) is set to 1e-6 throughout the study, unless specified otherwise. In the following, we refer to `strong coupling' when the solution is obtained by applying the fixed-point iteration method, whilst to `weak coupling' when fluid and solid equations are solved sequentially, but there is no iterative procedure for updating the luminal area. In this study, the entire model was implemented in Python 3.8, and all the simulation results were obtained using a desktop workstation with an Intel(R) Core(TM) i7-9700K CPU @ 3.60GHz.

\subsection{Assumptions on time constants}
The behaviour of the vascular contractility model was investigated at a steady state in our previous work~\cite{coccarelli2023} under different conditions (control and selective pharmacological modulation). By considering a cannulated vessel (no flow, same pressure at the extremities) immersed in a bath of physiological saline solution (PSS), here we aim to characterize its time-dependent response to upstream pressure variations. 
The dynamics of the SMC signalling are governed by the time constants ($\tau_{\text{c}0}$, $\tau_{\text{c}1}$, $\tau_{\text{c}2}$, $\tau_{\text{c}3}$, $\tau_{\text{c}4}$, $\tau_{\text{c}5}$, $\tau_{\text{c}6}$), each of them reflecting the speed of variation of an intracellular process. Reference experimental studies~\cite{johnson2009,moreno-dominguez2014} reported the diameter-vs-time recordings of different pressurized arteries upon the addition of ROCK and PKC inhibitors to the PSS bath. We observed that the duration of transient significantly differed across the reported dataset, and the number of traces may be too small to fully characterize the dynamics of the processes. Furthermore, the diffusion kinetics of the drug across the bath represents another source of uncertainty. In this study, we simply assume that $\tau_{\text{c}0} = \tau_{\text{c}1} = \tau_{\text{c}2} = \tau_{\text{c}}$ whilst all the other intracellular processes take place instantaneously ($\tau_{\text{c}3}$, $\tau_{\text{c}4}$, $\tau_{\text{c}5}$, $\tau_{\text{c}6}$ are assumed to be infinitesimal). Therefore, the total time required by the individual pressure-induced pathway (Ca$^{2+}$, ROCK, PKC) to convert the mechanical stimulus (pressure) into a signal for the contractile apparatus (actin-myosin filaments and cytoskeleton) is the same. For this time constant we initially considered different values $\tau_{\text{c}}$ = 1, 5, 10, 30, 60 s. The dynamics governing the actin-myosin filaments sliding within the passive surrounding matrix is characterized by the time constant $\tau_{\text{m}}$, which was estimated in the study by Murtada et al.~\cite{murtada2016} for mouse descending aorta ($\bar{\tau}_{\text{m}}$ = 6.188e-5 s). Due to the different vessel size and functionality between the latter and cerebral arteries, we also considered alternative values around the one proposed by the previous study (${\tau}_{\text{m}}$ = 1e3$\bar{\tau}_{\text{m}}$, 1e2$\bar{\tau}_{\text{m}}$, 1e1$\bar{\tau}_{\text{m}}$, $\bar{\tau}_{\text{m}}$, 1e-1$\bar{\tau}_{\text{m}}$). In all numerical experiments conducted in this work, the intracellular variables are initialized to 0, except for $\bar{u}_{\text{fs}}$ which is set to -2e-2, whilst $\lambda_{\theta}$ is set to 1.

\section{Results}
\subsection{Pressure-induced arterial wall dynamics}\label{res1}
\subsubsection{Comparison vs experimental protocols}
The dynamical behaviour of the vascular structure is assessed by simulating its response to alternated luminal pressure levels (between 10 and 60 mmHg) under control (PSS) and pharmacological modulation of the Ca$^{2+}$ pathway (via the addition of 30 $\mu$M Diltiazem to the PSS).
The effect of ${\tau}_{\text{c}}$ on all the modelled intracellular quantities upon pressure-activation is reported in the Appendix. Pressure variations within the 10 - 60 mmHg range induce a remarkable change in wall Ca$^{2+}$ concentration, while variations in ROCK activity and consequent MLCP phosphorylation are limited. This is in line with the hypothesis for which some Ca$^{2+}$ sensitization mechanisms are mainly activated at medium-high pressure levels ($>$ 60 mmHg)~\cite{osol2002}. On the other hand, it is difficult to assess and verify the simulated dynamics of the PKC pathway due to its unresolved bidirectional dependency with Ca$^{2+}$ activity in cerebral SMCs~\cite{osol1991,gokina1999,earley2007,el-yazbi2015}. Model predictions are compared against the experimental recordings by Knot and Nelson~\cite{knot1998}. In this numerical experiment, we only use the vascular mechanics model (described in Section \ref{msvm}) as the wall deformation and relative filament sliding are calculated by prescribing the luminal pressure as a function of time (see the Appendix for more details on the signal) and no `averaged active stress' is adopted. From~\cite{knot1998}, we assume the load-free ($P \approx$ 0 mmHg) outer diameter for the vessels in the control and inhibited case equal to 135 and 150 $\mu$m, respectively. All the other parameters of the wall mechanics model remain the same as in our previous study~\cite{coccarelli2023} (here and in the following sections), unless specified otherwise. For this simulation, the time step is set to 0.5 s. Diltiazem at a concentration of 30 $\mu$M is expected to prevent the opening of L-type calcium channels upon pressure-induced membrane depolarization, restricting the Ca$^{2+}$ influx into the intracellular space, which causes a reduction in tone development. Here, we assume that, upon drug effect, pressure has a limited effect on the intracellular Ca$^{2+}$ concentration, with only a marginal increase due to stretch-operated channels (see the Appendix for more details on the adopted pressure - Ca$^{2+}$ relationship). By considering different combinations of $\tau_{\text{c}}$ and $\tau_{\text{m}}$ (as described above), we estimate the associated RMSRE between the simulated and experimental outer diameter traces under control condition. The combination of parameters (${\tau}_{\text{c}}$ = 10 s, ${\tau}_{\text{m}}$ = 1e3$\bar{\tau}_{\text{m}}$ = 6.188e-2 s) yields the smallest discrepancy between simulated and experimentally-recorded outer diameter (Table \ref{tab1}).
\begin{table}[h!]
    \centering
    \begin{tabular}{|c|c|c|c|c|c|}
    \hline
  ${\tau}_{\text{c}}$ (s) & $\tau_{\mathrm{m}}$ = 1e3$\bar{\tau}_{\mathrm{m}}$  
  & $\tau_{\mathrm{m}}$ = 1e2$\bar{\tau}_{\mathrm{m}}$
  & $\tau_{\mathrm{m}}$ = 1e1$\bar{\tau}_{\mathrm{m}}$
  & $\tau_{\mathrm{m}}$ =    $\bar{\tau}_{\mathrm{m}}$
  & $\tau_{\mathrm{m}}$ = 1e-1$\bar{\tau}_{\mathrm{m}}$\\ \hline
 1 &1.657e-01&1.682e-01&1.561e-01&1.539e-01&1.537e-01\\
5 &1.460e-01&1.464e-01&1.442e-01&1.439e-01&1.438e-01\\
10 &1.367e-01&1.430e-01&1.454e-01&1.454e-01&1.453e-01\\
30 &1.592e-01&1.646e-01&1.665e-01&1.665e-01&1.665e-01\\
60 &1.807e-01&1.853e-01&1.863e-01&1.864e-01&1.864e-01\\ \hline
    \end{tabular}
\caption{Root mean square relative error (RMSRE) between simulated and experimentally recorded outer diameter upon control condition (from~\cite{knot1998}) across the ${\tau}_{\text{c}}$ - ${\tau}_{\text{m}}$ parametric space. For the calculation of the error we considered data points between 14.1 and 26.4 min.}
\label{tab1}
\end{table}
% Error mat [[0.16574518 0.16802765 0.15612853 0.15389723 0.15366793]
%  [0.14603245 0.1464431  0.14418097 0.1438773  0.14381177]
%  [0.13671106 0.14299094 0.14541482 0.14536252 0.14533745]
%  [0.15917215 0.164589   0.16646358 0.16652978 0.16653146]
%  [0.18073043 0.18526293 0.18633307 0.18642102 0.18642878]]
The RMSRE seems to be more sensitive to changes in the intracellular signalling time constant ${\tau}_{\text{c}}$, rather than to the time constant associated with the actin-filaments sliding ${\tau}_{\text{m}}$. Fig. \ref{val} shows how the intracellular Ca$^{2+}$ concentration and the outer diameter change in cerebral arteries when alternated luminal pressure levels are applied for the control and inhibition (30 $\mu$M Diltiazem) cases.
\begin{figure}
\centering
\includegraphics[width=1\linewidth]{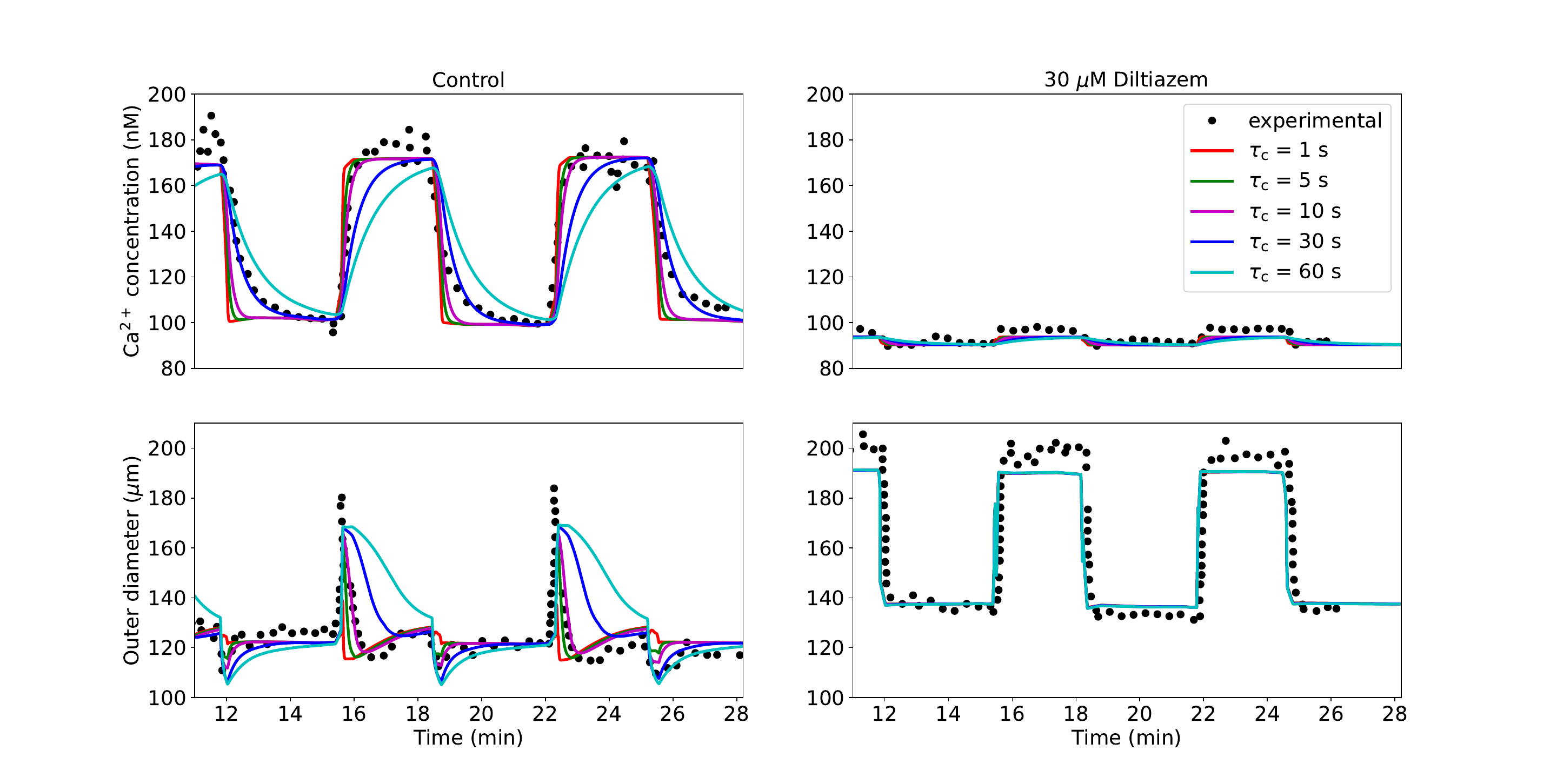}
\caption{Ca$^{2+}$ and outer diameter ($d_{\text{o}}$) responses to variable luminal pressure level in rat cerebral arteries for different time constants ${\tau}_{\text{c}}$. Responses are reported upon control and Ca$^{2+}$ modulation (30 $\mu$M Diltiazem) conditions. For the simulated curves, $\tau_{\mathrm{m}}$ = 1e3$\bar{\tau}_{\mathrm{m}}$. The maximum Ca$^{2+}$ intracellular concentration is set to 230 nM~\cite{cole2011}. The reported experimental recordings, as well as the associated luminal pressure time-dependent signals, are taken from the study by Knot and Nelson~\cite{knot1998}.}
\label{val}
\end{figure}
As expected, the timescale of the pressure-induced intracellular processes ($\tau_{\text{c}}$) plays a profound effect on the shape of the cytosolic Ca$^{2+}$ concentration as well as on the outer diameter responses. 
In the control case, a smaller ${\tau}_{\text{c}}$ corresponds to a higher Ca$^{2+}$ increase/decrease rate, while larger time constants are associated with a significantly slower variation in Ca$^{2+}$. Although
${\tau}_{\text{c}}$ = 30 s closely matches the Ca$^{2+}$ drop after pressure lowering, smaller time constants better represent the pressure-induced Ca$^{2+}$ elevation. The disparity in optimal ${\tau}_{\text{c}}$ values between the pressure-induced Ca$^{2+}$ increase and decrease may be due to the different Ca$^{2+}$ dynamics associated with each phase. The effect of different combinations ${\tau}_{\text{c}}$ - ${\tau}_{\text{m}}$ on the intracellular Ca$^{2+}$ concentration and outer diameter responses in the control case is reported in the Appendix.
Overall, ${\tau}_{\text{c}}$ = 10 s represents the best choice for capturing the experimental Ca$^{2+}$ trace over time. The same can be said regarding the diameter time evolution in the control case, as ${\tau}_{\text{c}}$ = 10 s provides the closest match to the corresponding experimental recording. 
30 $\mu$M Diltiazem is expected to inhibit the main Ca$^{2+}$ influx due to pressure, and therefore the impact of $\tau_{\text{c}}$ remains extremely marginal for both simulated responses (Ca$^{2+}$ and $d_{\text{o}}$). In this case, although the predicted diameter upon pressurization remains slightly underestimated, its rate of change (both increase and decrease) is accurately captured by the proposed model. 
%This simple analysis indicates that $\tau_{\text{c}}$ around 10 s seems the most appropriate choice for capturing the considered experimental data. On the other hand, $\bar{\tau}_{\text{m}}$ influences the pace at which filament sliding changes upon pressure variation but plays a very little role in the tissue (diameter) response.

\subsection{Pressure-induced dynamics across an arterial network}
The effect of myogenic tone on blood flow regulation can be well appreciated by evaluating haemodynamic quantities across a cerebral arterial network. Since the focus of this work is on the definition of a suitable methodology for modelling blood flow within self-regulated vessels, we consider an idealized symmetrical network branching from a rat's middle cerebral artery (reported in the previous section). Here, we introduce a vascular network (morphology reported in Fig. \ref{network}}) to assess i) the accuracy and efficiency of the different solution procedures and ii) the effect of upstream pressure changes on the system's blood flow dynamics. The generation $G$3 represents the last artery preceding the arteriolar vasculature in the parenchymal space. Parenchymal arterioles also develop myogenic tone, and they are expected to play an important role in the stabilization of flow and perfusion pressure in the face of upstream pressure changes~\cite{iadecola2017}. However, their structure and intracellular signalling factors may differ from cerebral arteries~\cite{cipolla2014b,li2017}. For the sake of simplicity, we do not explicitly represent these vascular beds (and their downstream vessels) and the network is truncated after the $G$3 vessels. The downstream circulation is represented through an (outlet) pressure boundary condition $P_{\text{out}}$, and backward pressure wave reflections are prevented by including a characteristic impedance $Z$ between the terminal node of each $G$3 vessel and the associated outlet node (where $P_{\text{out}}$ is imposed). To reflect in vivo conditions, $P_{\text{out}}$ and $P_{\text{ext}}$ are set to 50 mmHg and 10 mmHg, respectively. 
\begin{figure}[h!]
\centering
\includegraphics[width=1\linewidth]{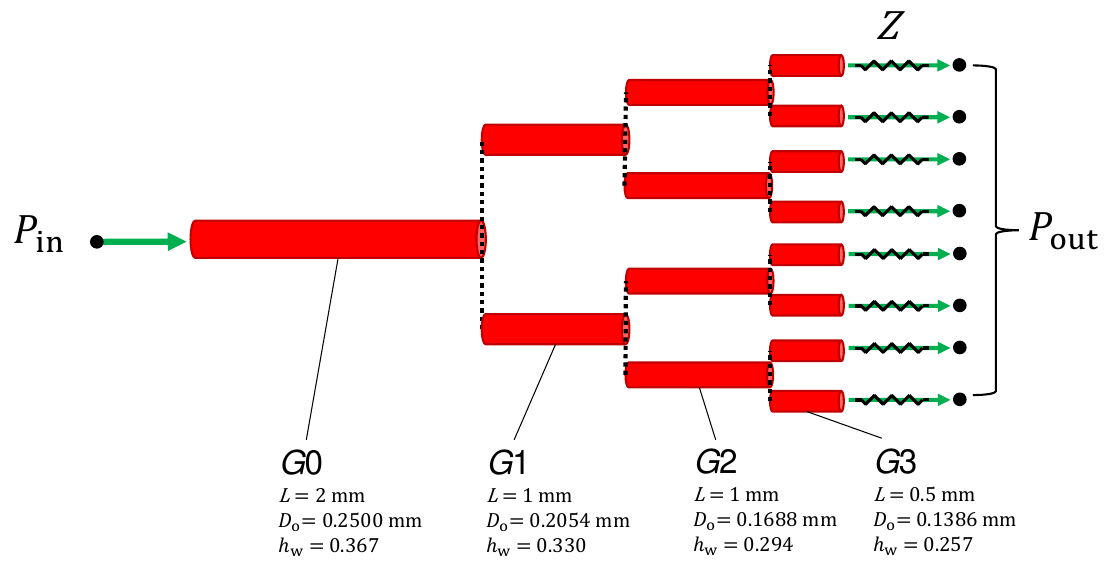}
\caption{Arterial network morphology. Vessel generations $G$0, $G$1 and $G$2 branch out into symmetrical branches and green arrows indicate the fluid flow direction at the network boundary nodes. $P_{\text{in}}$ and $P_{\text{out}}$ are, respectively, the pressures set at the inlet and outlet of the network, while $Z$ is the characteristic impedance associated with the terminal vessel. $L$ is the stretched vessel length (accounting for $\lambda_{z}$), $D_\text{o}$ is the load-free outer diameter, while $h_{\text{w}}$ is the ratio between thickness and mean radius under load-free conditions. The load-free outer diameters for the generations $G$1, $G$2 and $G$3 are derived from experimentally measured branching patterns of the cerebral arterial tree (the area-ratio between parent and daughter vessels is set equal to 1.35)~\cite{helthuis2019}. To account for the gradual decrease in wall thickness along the tree, $h_{\text{w}}$ for $G$1, $G$2 and $G$3 are assumed to be, respectively, 90$\%$, 80$\%$ and 70$\%$ of the $G$0 value.  
%The size of the vessels is chosen in line with experimental studies on rat pial circulation~\cite{johnson2009}.
}
\label{network}
\end{figure}
Blood viscosity and density are set to 0.05 poise and 1.04 g$\cdot$cm$^{-3}$, respectively. In each vessel, the axial spatial domain is discretized with two elements. Here and in the following the pressure is initialized with the first inlet value whilst the initial flow rate is set to 0 ml/s.

\subsubsection{Comparison between solution procedures}\label{res2}
Here, we evaluate the impact of numerical procedure settings such as the time step $\Delta t$, coupling type and active stress averaging on the accuracy and efficiency of the solution across the network. To appropriately test the model robustness, we prescribe a periodic pressure signal with a mean that varies over time at the inlet. We consider the simulation results obtained with $\Delta t$ = 1e-4 s, strong coupling and without averaged active stress as the ground truth solution (case 1).
\begin{figure}[h!]
\centering
\includegraphics[width=1\linewidth]{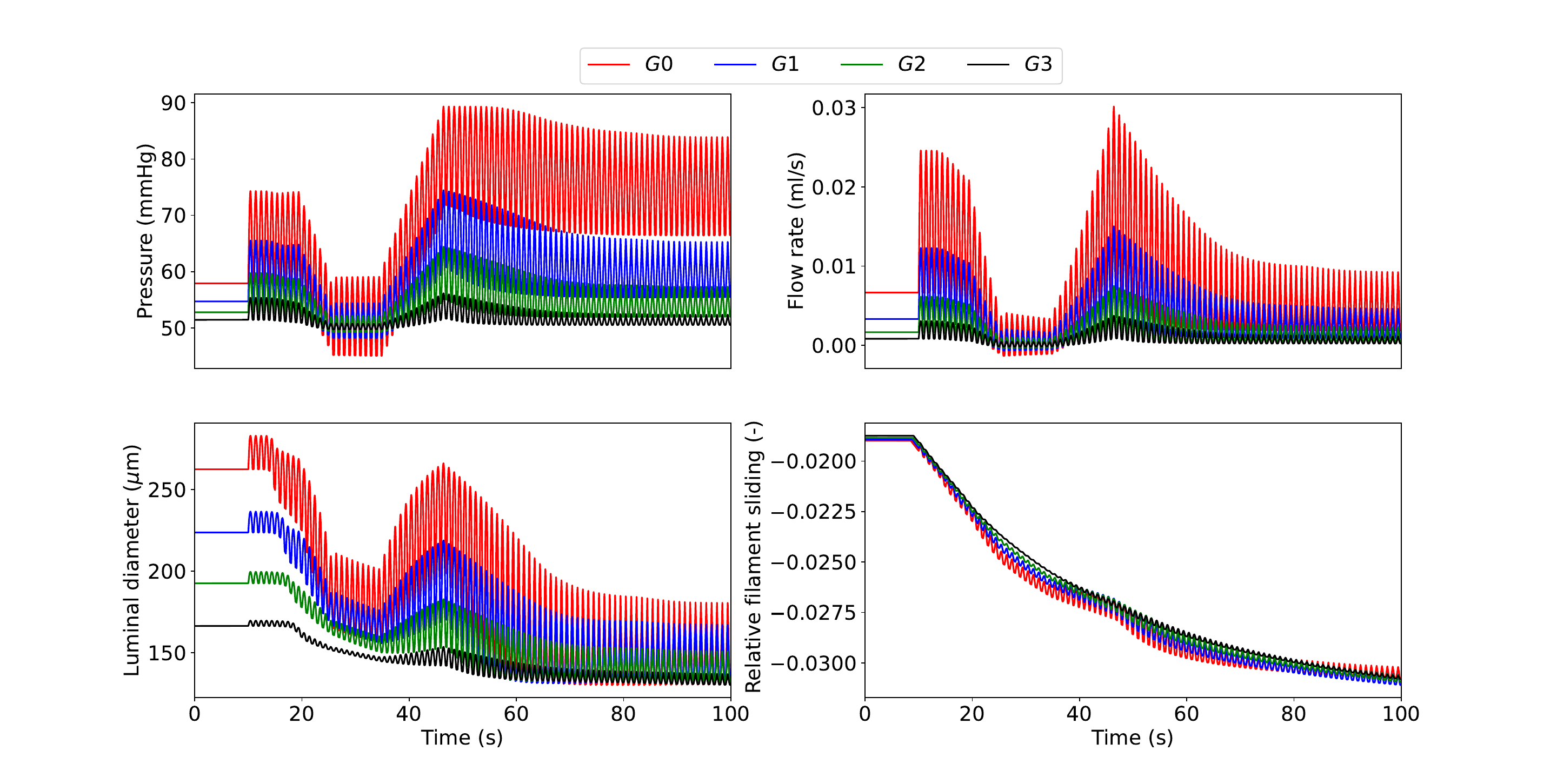}
\caption{Luminal pressure ($P$), flow rate ($Q$), luminal diameter ($d_{\text{i}}$) and relative filament sliding ($\bar{u}_{\textbf{fs}}$) vs time upon variable upstream pressure across the vascular network. All quantities are reported at the midpoint of the axial length of the associated generation vessel. The time-dependent pressure signal at the network inlet is included in the Appendix.}
\label{sinsig}
\end{figure}
Dramatic changes in upstream pressure are accompanied by substantial variations in flow rate, luminal diameter and actin-myosin filament sliding across the vessel network (Fig. \ref{sinsig}). As expected, the fluctuation amplitude in all the recorded variables is significantly more mitigated in the higher-generation vessels than in the upstream larger arteries. Alternatively, more computationally efficient simulation settings than case 1 are explored (Table \ref{cases}). The numerical accuracy of cases 2-5 was assessed (against case 1) by evaluating how RMSRE for flow ($Q$) and luminal area ($A$) distribute across the vascular network (Table \ref{RMSRE}).
\begin{table}[h!]
    \centering
    \begin{tabular}{|c|c|c|c|}
    \hline
    Case & \makecell{Time step \\ $\Delta t$ (s)} & Coupling & \makecell{Averaged \\ active stress} \\\hline
    1 & 1e-4 & strong & no \\ \hline
    2 & 2e-4 & strong & no \\ \hline
    3 & 2.5e-4 & strong & no \\ \hline 
    4 & 2.5e-4 & weak & no\\ \hline  
    5 & 2.5e-4 & weak & yes \\ \hline
    \end{tabular}
    \caption{Considered simulation settings.}
        \label{cases}
\end{table}

\begin{figure}[h!]
\centering
\includegraphics[width=1\linewidth]{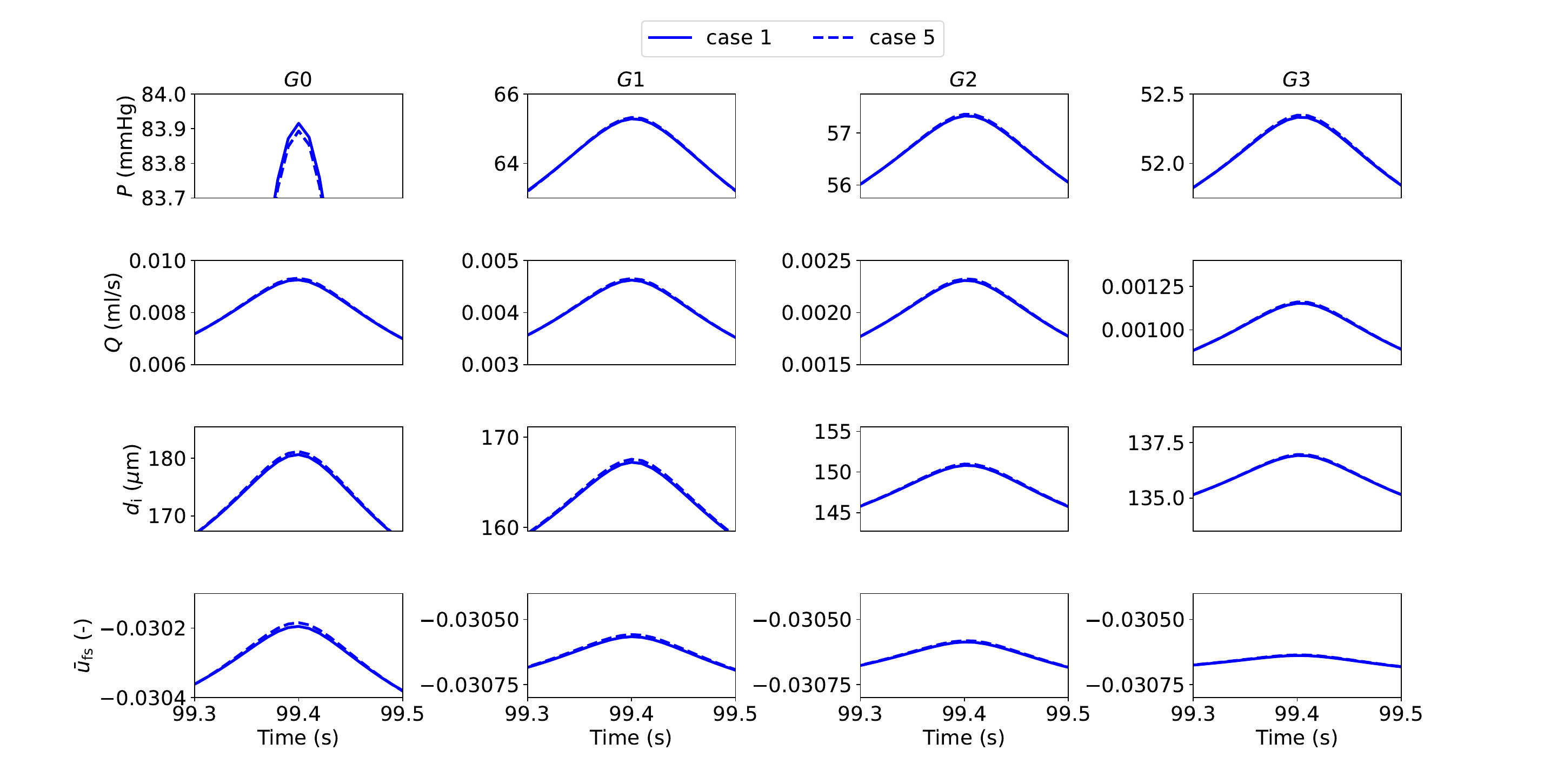}
\caption{Details on luminal pressure ($P$), flow rate ($Q$), luminal diameter ($d_{\text{i}}$) and relative filament sliding ($\bar{u}_{\textbf{fs}}$) vs time across the vascular network for cases 1 and 5. The reported variables are recorded at the middle point of the axial length of the associated generation vessel. The comparison is carried out by considering simulation solutions recorded every 1e-2
s.}
\label{diff}
\end{figure}

\begin{table}[h!]
    \centering
    \begin{tabular}{|c|c|c|c|c|c|c|c|}
    \hline
    Case $\#$ &\makecell{Relative\\ WCT (\%)}& \makecell{Mean $Q$\\ RMSRE (-)} & \makecell{Min $Q$\\ RMSRE (-)}& \makecell{Max $Q$\\ RMSRE (-)} & \makecell{Mean $A$ \\ RMSRE (-)}& \makecell{Min $A$ \\ RMSRE (-)}& \makecell{Max $A$ \\ RMSRE (-)} \\ \hline
    2 &57.54& 1.110e-2 &  3.01e-3 ($G0$) & 2.699e-2 ($G$2) &9.1e-6 &6.3e-6 ($G$0) & 1.22e-5 ($G3$) \\ \hline 
    3 & 53.35 & 1.665e-2 & 4.51e-3 ($G$0)&4.048e-2 ($G$2) & 1.37e-5& 9.7e-6 ($G$0) & 1.82e-05 ($G$3)
    \\ \hline   
    4 & 20.56 & 1.661e-2  &4.48e-3 ($G$0)& 4.036-2 ($G$2)& 9.43e-5&4.07e-5 ($G$3) & 1.312e-4 ($G$1)  \\ \hline    
    5 & 11.62& 1.831e-2 & 7.65e-3 ($G$0)& 4.048e-2 ($G$2) &1.5278e-3&2.968e-4 ($G$3) & 3.8427e-3 ($G$0)    \\ \hline
    \end{tabular}
\caption{Relative Wall Clock Time (WCT) and accuracy for the considered numerical settings (with respect to case 1). The wall clock time for case 1 was 2.73608e5 s. For each arterial generation, the flow $Q$ (or area $A$) RMSRE is evaluated at the middle of the axial length of the vessel. The mean flow $Q$ (or area $A$) RMSRE is obtained by averaging the values across all the vessel generations of the network. Min and Max values are reported together with the associated vessel generation (in brackets). The comparison is carried out by considering simulation solutions recorded every 1e-2 s.}\label{RMSRE}
\end{table}
%case2: 157426.219s= 1.5742622e5s
%case3: 145969.188s=1.45969188e5s (old 109676.746s=1.09676746e5s)
%case4: 56265.949s=5.6265949e4s(old 32058.143s=3.2058143e4s)
%case 5:31797.655s=3.1797655se4s(old 19499.616s=1.9499616e4s)
Simulation results indicate that a time step equal to 2.5e-4 s provides an optimal compromise between numerical accuracy and computational speed. Weak coupling provides another significant reduction in Wall Clock Time (WCT) without remarkably affecting the solution precision (all $Q$ RMSREs in case 4 are actually slightly improved vs case 3). This information may be extremely important when the model is used to describe vast vascular networks and/or simulate more comprehensive cellular dynamics. Adopting an `averaged active stress' enables a further significant reduction in computational time and maintains the error below a reasonable threshold (Mean $Q$ RMSRE $<$ 1.9 $\%$, Mean $A$ RMSRE $<$ 0.16 $\%$). The discrepancy between the numerical solutions obtained with the most and least computationally demanding strategies (cases 1 and 5) is shown in Fig. \ref{diff}.
The time-evolution of the relative error for flow rate and luminal area of cases 4 and 5 (with respect to case 1) across the vascular network indicate that in both cases the accuracy of the solution does not significantly deteriorate over the considered time interval (see Appendix for the error comparison between cases).
The numerical strategy associated with case 5 is adopted for the simulations in the next section. 
% \begin{table}[h!]
%     \centering
%     \caption{Accuracy and computational time for different numerical strategies. The (network) flow rate RMSRE is obtained as the average between the errors calculated at the middle of each vessel generation.}
%     \label{networkData}
%     \begin{tabular}{|c|c|c|c|c|c|c|c|}
%     \hline
%     Case & \makecell{Time step \\ $\Delta t$ (s)} & Coupling & \makecell{Averaged \\ active stress} & \makecell{Flow \\ RMSRE (-)} &  \makecell{Area \\ RMSRE (-)} & \makecell{Computational \\ time (s)} \\ \hline
%     1 & 2e-4 & Strong & No & - & - & 2.129526e5 \\ \hline
%     2 & 2.5e-4 & Strong & No & 2.78e-4 & & 1.850914e5 \\ \hline
%     3 & 4e-4 & Strong & No & 1.11e-3 & & 1.430068e5 \\ \hline 
%     4 & 4e-4 & Weak & No & 7.26e-4 & & 4.52239e4 \\ \hline  
%     5 & 4e-4 & Weak & Yes & 1.15e-3 & & 2.61185e4 \\ \hline
%     \end{tabular}
% \end{table}

\subsubsection{Myogenic response to upstream pressure increase}\label{res3}
\begin{figure}[h!]
\centering
\includegraphics[width=1\linewidth]{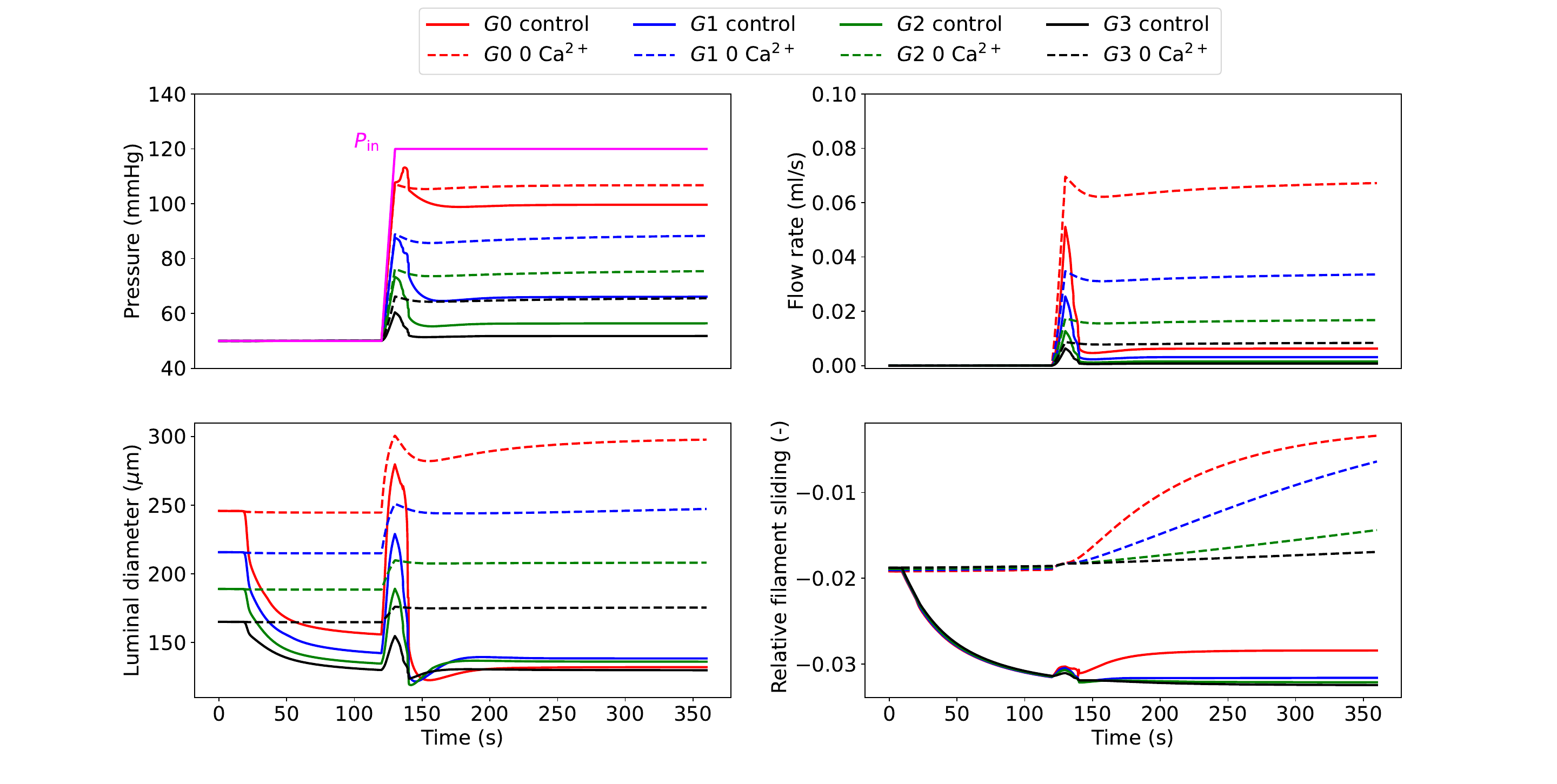}
\caption{Luminal pressure ($P$), flow rate ($Q$), luminal diameter ($d_{\text{i}}$) and relative filament sliding ($\bar{u}_{\textbf{fs}}$) vs time upon upstream pressure change (ramp from 50 to 120 mmHg) across the vascular network. The reported variables are recorded at the middle point of the axial length of the associated generation vessel. The time-dependent pressure signal at the network inlet ($P_{\mathrm{in}}$) is depicted with a solid magenta line.}
\label{ramp}
\end{figure}

\begin{figure}[h!]
\centering
\includegraphics[width=1\linewidth]{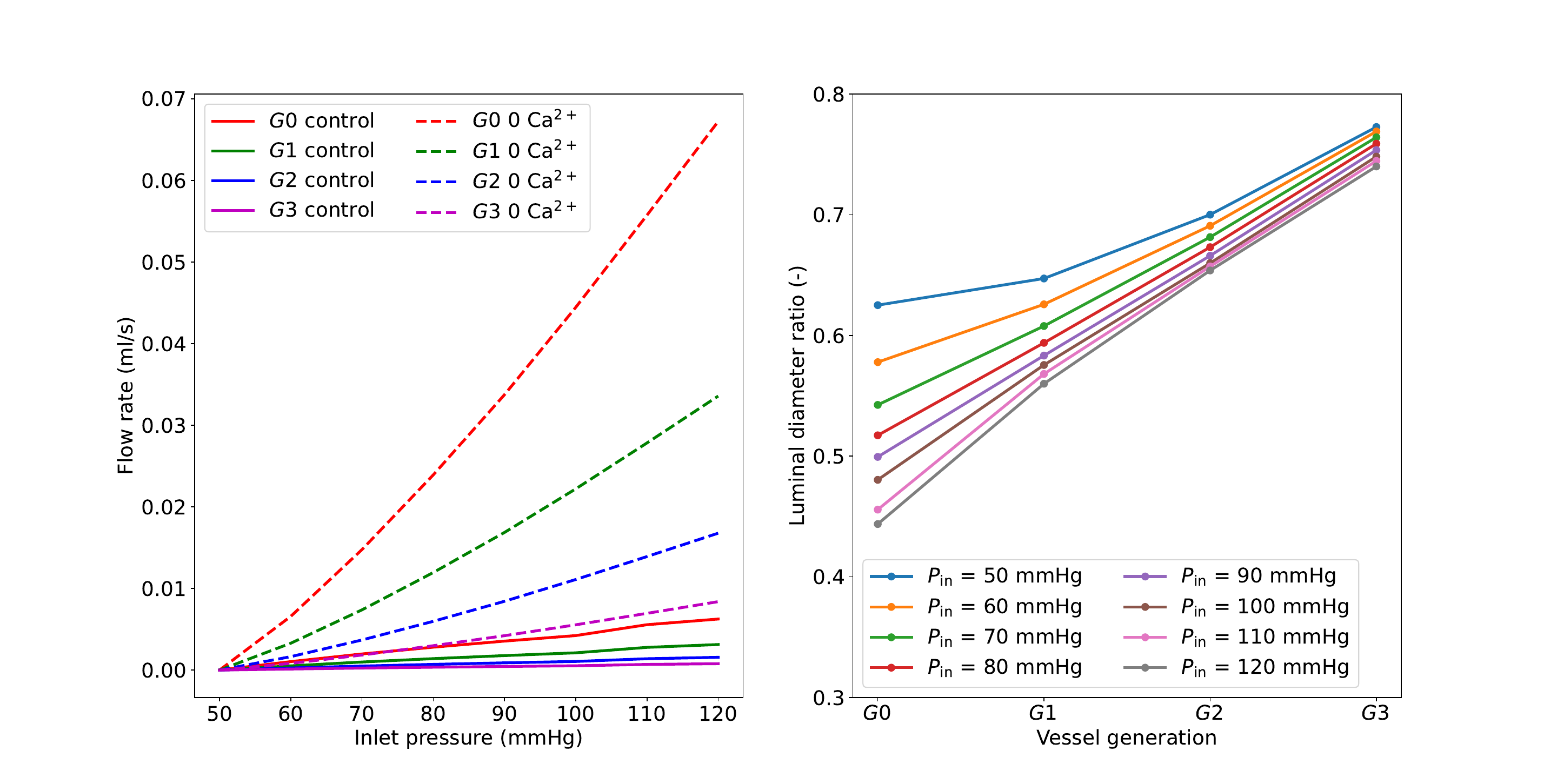}
\caption{Effect of upstream pressure on flow rate and luminal diameter ratio (control over 0 Ca$^{2+}$) distribution across vessel network. The reported simulated data are at steady state conditions (at 360 s), and the flow rate is recorded at vessel $G$0.  The reported quantities are recorded at the middle point of the axial length of the associated generation vessel. The inlet pressure is elevated from 50 mmHg to the final pressure by using a linear ramp between 120 - 130 s.}
\label{presFlowCurve}
\end{figure}

When the upstream pressure increases, the myogenic tone enables small arteries (and arterioles) to adjust their luminal diameters to stabilize blood flow and limit perfusion pressure variations. Fig. \ref{ramp} shows the myogenic response across all the vessel generations to an extreme upstream pressure change from 50 to 120 mmHg.
The pressure surge initially causes an increase in diameter and flow, which are then gradually reduced as the tone develops. The propagated pressure change from upstream is enormously mitigated at higher-generation vessels. Alongside the curves representing myogenically active arteries (control), we also reported the predictions for vessels with impaired contractile capacities due to extracellular Ca$^{2+}$ removal (0 Ca$^{2+}$). The comparison between these vessel conditions highlights the importance of myogenic tone in counteracting acute hydrodynamic changes. Under control conditions, all vessels constrict to redistribute the new pressure load and minimize flow variations, while in the presence of tone (partial) inhibition, the vessels dilate upon pressure increase. The variation of $\bar{u}_{\textbf{fs}}$ in time shows how functioning contractile units respond to a pressure surge.
We evaluated the impact of the final (steady-state) inlet pressure level on the total flow rate through the vascular system (recorded at $G$0). The steady-state results obtained in the control and 0 Ca$^{2+}$ cases define two distinct relationships between upstream pressure and flow (Fig. \ref{presFlowCurve}). Upon inlet pressure increase, control conditions are associated with moderate flow rate increments, whilst 0 Ca$^{2+}$ case exhibits almost an exponential dependency between flow rate and inlet pressure.
The luminal diameter ratio (control over 0 Ca$^{2+}$) is used to define the level of vessel constriction across the network for various final upstream pressures. For the boundary conditions considered, large myogenically active vessels are subjected to higher luminal pressure and provide more resistance to flow than the smaller arteries. Although the proposed numerical experiments allow us to quantify the impact of myogenic tone on blood flow stabilization across a network of (rat) cerebral arteries, the associated results may vary depending on the network morphology and role of the downstream circulation.

\section{Discussion}
Thanks to myogenic tone, cerebral arteries and arterioles can locally adjust their diameter to maintain nearly constant blood flow and perfusion pressure despite acute changes in upstream pressure. Therefore, this local blood flow control helps prevent eventual local tissue oxygen and nutrient starvation. Malfunctioning of this regulatory mechanism can contribute to the development and progression of different life-threatening conditions, and indeed, its restoration has been proposed as a potential therapeutic target~\cite{hill2009,palomares2014,el-yazbi2017,lidington2018}. Despite its physiological importance and implication in diverse pathological conditions, the effect of myogenic tone on blood flow dynamics in cerebral vessels has been rarely analyzed and quantified through in-silico approaches. The few currently available models, although computationally-efficient, do not directly relate the intracellular processes underlying myogenic tone to blood flow changes across vascular networks. Therefore, there is an urgent need for more physiologically-representative modelling methodologies which are capable of dealing with different types of data from the laboratory and clinic \cite{payne2024}.

Through this work, we propose a comprehensive multi-scale computational methodology which integrates, in a robust manner, the vascular SMC contractile machinery into blood flow dynamics in cerebral arteries. This framework aims to complement, confirm and conceive future investigations on blood flow regulation due to myogenic tone across the cerebral circulation. The time-dependent behaviour of the model was investigated at both single vessel and network levels. The comparison against the experimental recordings of the vessel wall Ca$^{2+}$ concentration and diameter allowed us to identify the time constants governing the response to luminal pressure variations. The results obtained with the simulated drug intervention (30 $\mu$M Diltiazem) further validated the proposed methodology. Altogether, these results proved that the developed model is able to capture the pressure-inducing wall dynamics under different conditions with a good level of accuracy (RMSRE $<$ 14 $\%$ with the identified time constants).

The main aim of this work was to assess the proposed fluid-structure interaction methodology in terms of numerical convergence and performance. To do this, we defined an idealized network made of a rat's middle cerebral arteries and its three (symmetric) generations. The simple morphology of the vascular network hindered the comparison against experimental flow measurements but allowed to clearly quantify the haemodynamics for each vessel category. We prescribed a highly variable signal for the pressure at the inlet whilst reflection-free conditions were imposed at the outlets of the vascular network.
By considering this newly defined toy-problem, we analyzed the impact of different numerical settings on the computational speed and accuracy of the solution. Weakly coupling and considering an average active stress allows for a substantial reduction in computational time, without significantly sacrificing the accuracy of the solution.
Imposing the reflection-free outflow conditions allows to evaluate exclusively the contribution of $G$0-$G$3 vessels to the network flow dynamics but requires a severe time step ($\Delta$t $\le$ 2.5e-4 s). Under such outflow conditions, simulations with $\Delta$t $>$ 2.5e-4 s (i.e., $\Delta$t = 5e-4 s) experienced convergence issues. The numerical instability behind this may be associated with the resulting vessel wave speed (of pulse propagation), which can be also influenced by the boundary conditions. We indeed observed that by using a three-element Windkessel model as an outflow boundary condition, this time step limitation was relaxed (results not shown). Overall, the numerical strategy of case 5 represents a good compromise between accuracy and computational costs for problems with a time span of a few minutes. For problems with longer time duration and where high accuracy is necessary, we recommend weak coupling without average active stress.

The model was then used to evaluate how a change in pressure at the inlet of a middle cerebral artery is mitigated (via myogenic mechanism) across an arterial network composed of its three generations. The computed results highlight the importance of myogenic tone in limiting blood flow variation upon significant pressure changes. The proposed framework is based on a biochemical model for the vascular wall~\cite{coccarelli2024} whose parameters were identified by considering both intracellular and tissue recordings under different conditions (control, no extracellular Ca$^{2+}$, with vasoactive agents). With these settings, the current framework predicts that any increase in upstream pressure will lead to a moderate rise in blood flow across the arterial network (compared to the case without extracellular Ca$^{2+}$). This might seem to slightly contrast with the `theoretical' autoregulation curve `flow rate - upstream pressure' (see~\cite{aletti2016a} for instance), for which flow remains constant across a mid-pressure interval. However, previous studies~\cite{bryan2001a,toth2011} demonstrated that intraluminal flow also contributes to tone development in cerebral arteries. In pressurized pial vessels, an increase in flow causes further vasoconstriction, and this seems to synergistically work together with the myogenic mechanism to preserve blood volume within the intracranial space~\cite{koller2012}. The current modelling framework does not account yet for the flow-induced tone regulation, and this may explain why the predicted flow rate is not maintained constant across the upstream pressure range 60 - 100 mmHg but presents a linear dependency which is in line with the data reported in~\cite{toth2011}. We are currently investigating this regulatory component.
To conclude, the presented framework represents an essential tool for investigating and quantifying blood flow regulation mechanisms in cerebral circulation. If provided with flow information at the inlet and outlet, the model can recover the mechanical stimuli acting on SMCs along the vascular network, as well as their contribution to blood flow regulation. While this model can already be used to mimick some ex-vivo scenarios, we plan to extend its predictive capacity by incorporating new modelling components (such as flow-induced tone and metabolic function) and validating them against new in-vivo data.

\section*{Acknowledgments}
\textit{A.C. and I.P. acknowledge the support by Swansea University through the College of Engineering Zienkiewicz/Centenary
scholarship. O.F.H was supported by the NIH National Heart, Lung, and Blood Institute (R01HL169681), the National Institute on Aging (R21AG082193), the National Institute of General Medical Sciences (P20GM135007), the Bloomfield Early Career Professorship in Cardiovascular Research, the Totman Medical Research Trust, the Cardiovascular Research Institute of Vermont, and a }\textit{grant (2024-338506) from the Chan Zuckerberg Initiative DAF, an advised fund of Silicon Valley Community Foundation.}

%\bibliography{refs}
\bibliographystyle{acm}

\appendix
\section*{Appendix}
\subsection*{A1: Upstream pressure signal}
% \begin{figure}[h!]
%             \centering
% \includegraphics[width=1\linewidth]{inlet2.pdf}
%             \caption{Pressure time-dependent signal at inlet used in Section \ref{res2}.}
%             \label{inlet2}
% \end{figure}

% \begin{figure}[h!]
%             \centering
% \includegraphics[width=1\linewidth]{inlet3.pdf}
%             \caption{Pressure time-dependent signals at inlet used in Section \ref{res3}.}
%             \label{inlet3}
% \end{figure}

%\begin{figure}[h!]
%    \centering
%    \begin{subfigure}[b]{\linewidth} % Width for the first image (full width)
%        \centering
%        \includegraphics[width=1\linewidth]{inlet2.pdf} % Adjust size as needed
      %  \caption{Pressure time-dependent signal at inlet used in Section \ref{res2}.}
%        \label{fig:subfig1}
%    \end{subfigure}
    % \vspace{0.5em} % Optional: Adds vertical space between the images
%    \begin{subfigure}[b]{\linewidth} % Width for the second image (full width)
%        \centering
%        \includegraphics[width=1\linewidth]{inlet3.pdf} % Adjust size as needed
   %     \caption{Pressure time-dependent signals at inlet used in Section \ref{res3}.}
%        \label{fig:subfig2}
%    \end{subfigure}
    
 %   \caption{Inlet boundary conditions used in Section 3.2. Top: pressure time-dependent signal at inlet used in Section \ref{res2}. Bottom: pressure time-dependent signals at inlet used in Section \ref{res3}.}
 %   \label{inlets}
%\end{figure}

\begin{figure}[h!]
\centering
\includegraphics[width=1\linewidth]{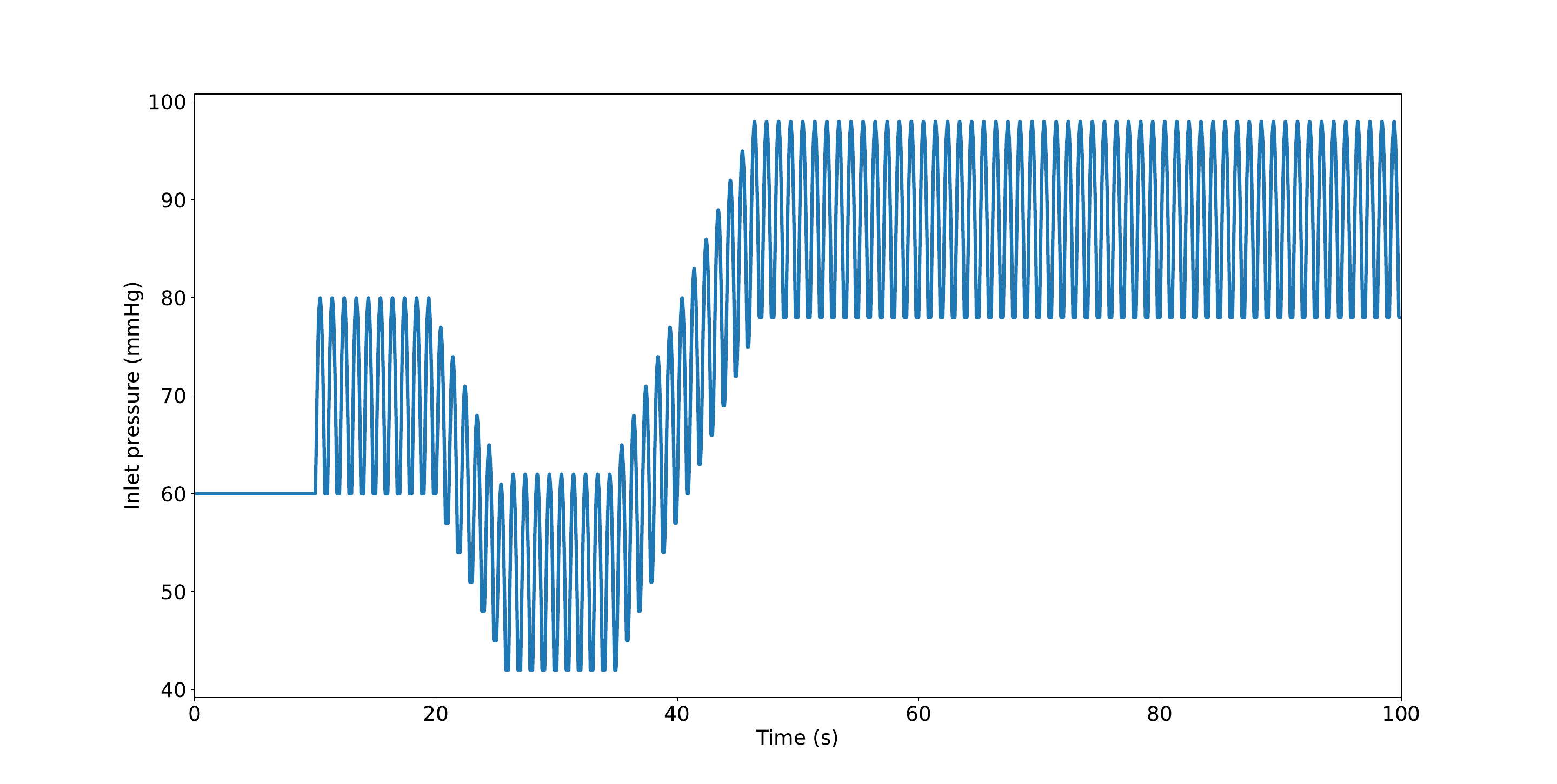} % Adjust size as needed
%  \caption{Pressure time-dependent signal at inlet used in Section \ref{res2}.}
\label{fig:subfig1}
% \vspace{0.5em} % Optional: Adds vertical space between the images
\caption{Pressure time-dependent signal at network inlet used in Section \ref{res2}.}
\label{inlets}
\end{figure}

\subsection*{A2: Effect of $\tau_{\text{c}}$ on intracellular quantities and tissue response to pressure}

\begin{figure}[h!]
\centering
\includegraphics[width=1.15\textwidth]{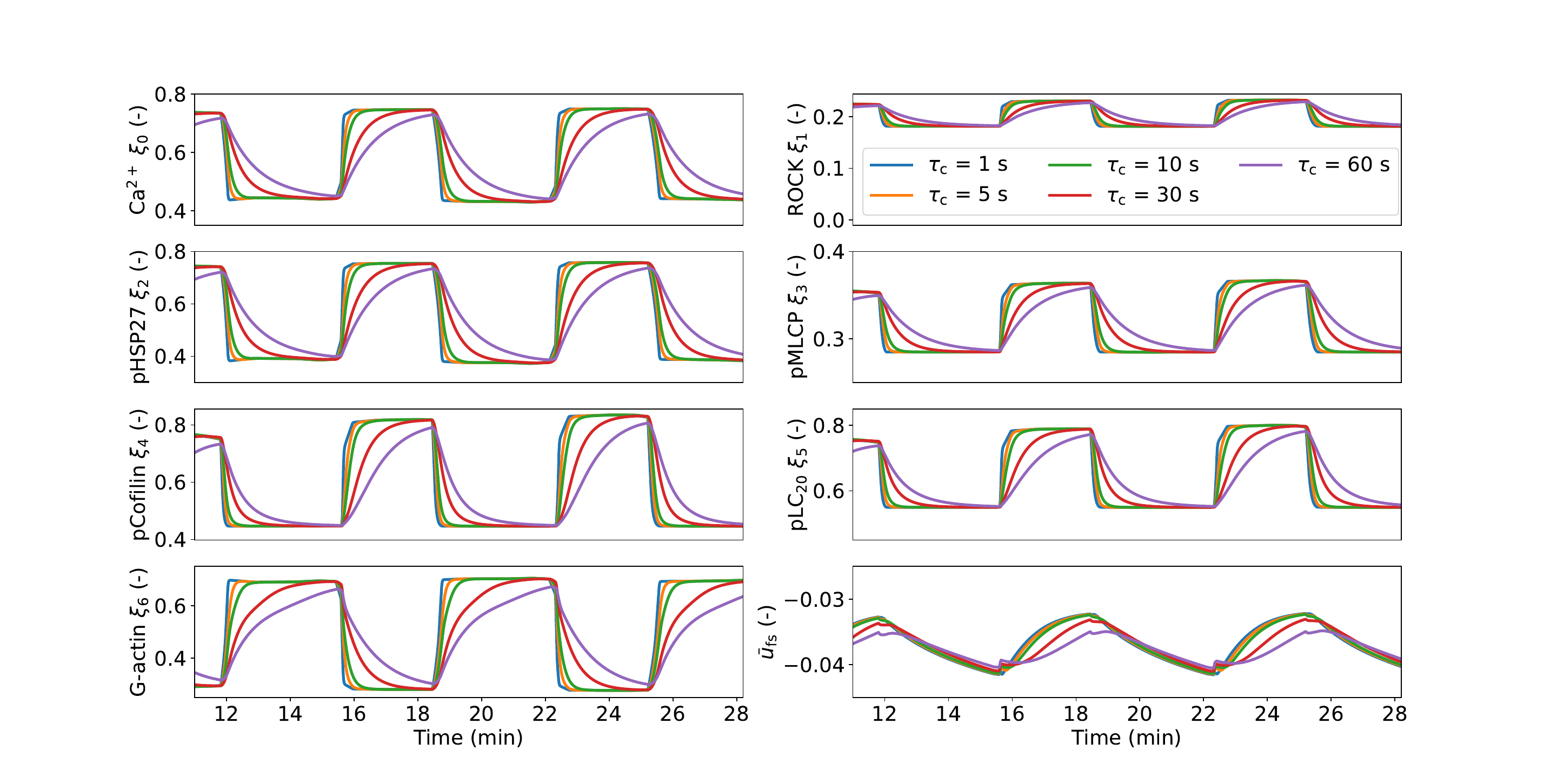}
\caption{Responses of the intracellular variables to changes in luminal pressure in rat cerebral arteries for different time constants $\tau_{\mathrm{c}}$.}
\label{intravar2}
\end{figure}

\begin{figure}[h!]
\centering
\includegraphics[width=1.15\textwidth]{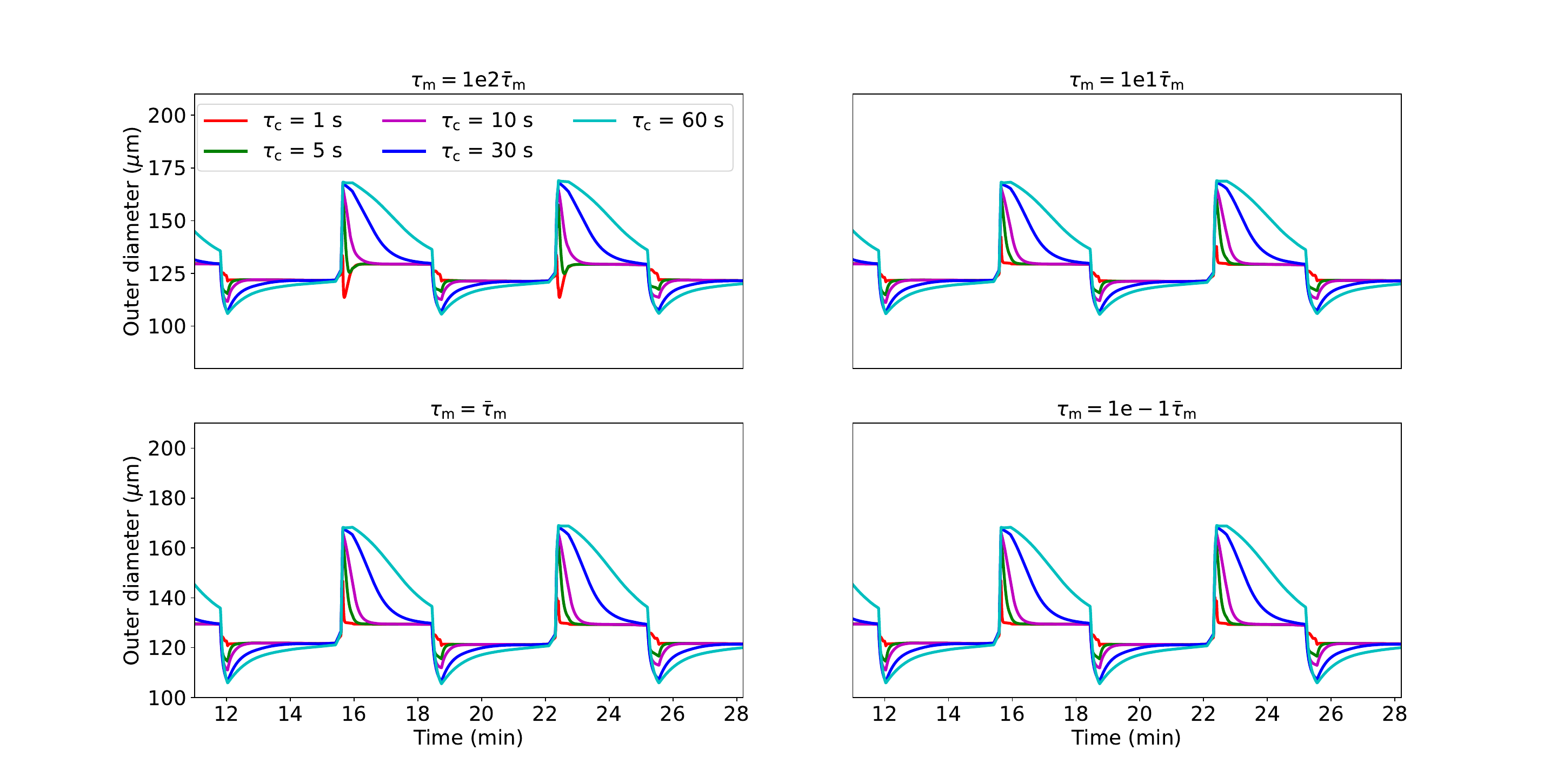}
\caption{Diameter response to changes in luminal pressure in rat cerebral arteries for different combinations of time constants $\tau_{\mathrm{c}}$ and $\tau_{\mathrm{m}}$. }
\label{tau_m}
\end{figure}

\subsection*{A3: Dependency of intracellular Ca$^{2+}$ on pressure}

\begin{figure}[h!]
\centering
\includegraphics[width=1.\linewidth]{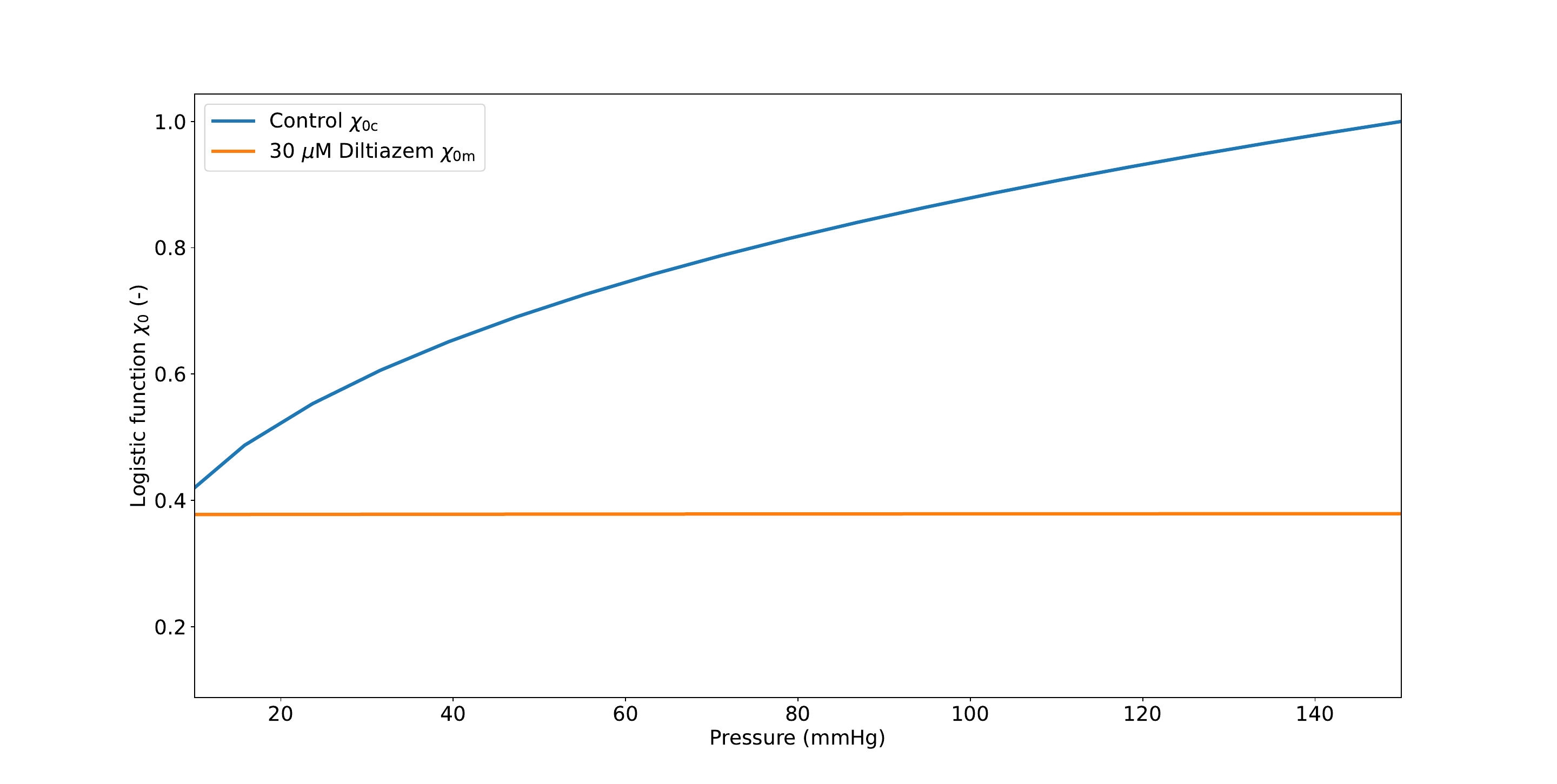}
\caption{Dependency of SMC intracellular Ca$^{2+}$ on luminal pressure for the control and Ca$^{2+}$ modulation cases. The logistic function associated with the Ca$^{2+}$ modulation case $\chi_{\mathrm{0m}}(P)$ is defined as a linear combination of the control $P$ - Ca$^{2+}$ curve $\chi_{\mathrm{0c}}(P)$: 
$\chi_{\mathrm{0m}}(P)$=$\chi_{\mathrm{0c}}(0.05)$ + $0.05[\chi_{\mathrm{0c}}(P)-\chi_{\mathrm{0c}}(0.05)]$.
}
\label{cacurves}
\end{figure}

\subsection*{A4: Time-dependency of relative errors for cases 4 and 5}

\begin{figure}[h!]
\centering
\includegraphics[width=1\linewidth]{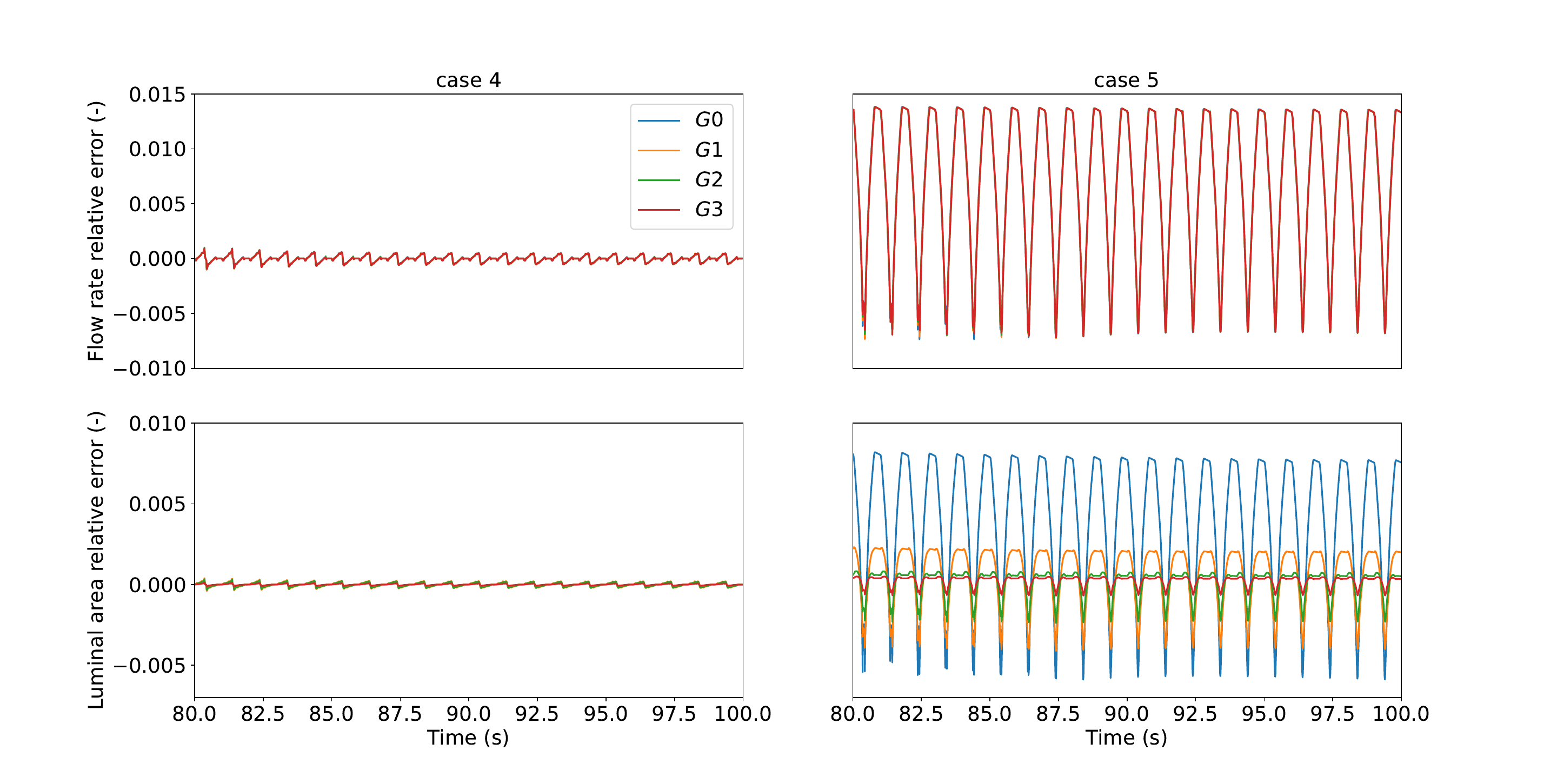}
\caption{Relative error in flow rate and luminal area vs time for cases 4 and 5 (with respect to case 1) across the vascular network. These are evaluated at the middle of the axial length of the vessel of each generation.}
\label{relerr}
\end{figure}

\end{document}